\documentclass{article}
\usepackage{epsfig,amssymb,float}
\input{epsf}
\newcommand{\be}{\begin{equation}}
\newcommand{\ee}{\end{equation}}
\newcommand{\ba}{\begin{eqnarray}}
\newcommand{\ea}{\end{eqnarray}}
\newcommand{\beq}{\begin{equation}}
\newcommand{\eeq}{\end{equation}}
\newcommand{\beqa}{\begin{eqnarray}}
\newcommand{\eeqa}{\end{eqnarray}}

\newcommand{\vs}{\vspace{-0.15cm}}
\begin{document}

\begin{flushright}
{\tiny  FZJ-IKP(TH)-2002-06} \\
{\tiny  $\,$} \\
\end{flushright}

\vspace{1cm}

\begin{center}

\bigskip

{{\Large\bf Novel analysis of chiral loop effects in the\\[0.3em]
generalized Gerasimov--Drell--Hearn sum rule
}}

\end{center}

\vspace{.3in}

\begin{center}
{\large
V\'eronique Bernard$^{\dagger,}$\footnote{email: bernard@lpt6.u-strasbg.fr},
Thomas R. Hemmert$^{\ast,}$\footnote{email: themmert@physik.tu-muenchen.de},
Ulf-G. Mei{\ss}ner$^{\ddagger,}$\footnote{email: u.meissner@fz-juelich.de}}

\vspace{1cm}

$^\dagger${\it Universit\'e Louis Pasteur, Laboratoire de Physique
               Th\'eorique\\
               F--67084 Strasbourg, France}

\bigskip

$^{\ast}${\it Technische Universit\"at M\"unchen, Physik Department T-39\\ 
              D-85747 Garching, Germany }

\bigskip

$^\ddagger${\it Forschungszentrum J\"ulich, Institut f\"ur Kernphysik
(Theorie)\\ D-52425 J\"ulich, Germany \\and\\
Karl-Franzens-Universit\"at Graz, Institut f\"ur Theoretische Physik\\
Universit\"atsplatz 5, A-8010 Graz, Austria}

\bigskip

\end{center}

\vspace{.6in}

\thispagestyle{empty}

\begin{abstract}\noindent
We study the chiral loop corrections to the generalized
Gerasimov-Drell-Hearn sum rule of the nucleon  for finite
photon virtuality in the framework of a Lorentz--invariant
formulation of baryon chiral perturbation theory. We 
perform a complete one--loop calculation and obtain significant
differences to previously found results based on
the heavy baryon approach for the proton and neutron 
spin--dependent forward Compton amplitudes.
\end{abstract}

\vfill

\pagebreak

\noindent {\bf 1.} The spin structure of the nucleon is under active
theoretical and experimental investigation. Of particular interest
is the transition from the perturbative regime at high momentum
transfer to the non--perturbative  low--energy region. 
Interestingly, systematic 
and controlled theoretical calculations are only available for very
small or very large momentum transfer while the intermediate region
is accessible to resonance models or can be investigated using
dispersion relations. Of special interest is the generalized  
Gerasimov-Drell-Hearn (GDH) sum rule \cite{GDH}
for virtual photons. It was pointed out in \cite{BKM} that one
can systematically investigate its momentum
evolution at low energies
in the framework of chiral perturbation theory (CHPT). In a series
of papers, Ji et al. \cite{Ji1,Ji2} proposed a formalism that is
better suited for the direct connection with the spin--dependent
structure functions measured in deep inelastic  scattering. A complete
one--loop fourth order calculation was presented for the forward
spin--dependent Compton amplitudes and it was shown that the 
$Q^2$--variation was much stronger than expected from naive dimensional
analysis. While the proton and the neutron spin--dependent amplitudes
receive important resonance contributions that can not be calculated
explicitly in chiral perturbation theory~\footnote{An exception to
  this is the contribution from the $\Delta (1232)$ which can be
 systematically analyzed in the framework of the small scale expansion
 \cite{HHK}. However, at present a relativistic formulation 
 for the $\Delta$ based on
 a systematic power counting has not been worked out.}
it was pointed out by Burkert \cite{VB} that in the
proton--neutron difference these resonance contaminations largely drop
out and also that the $Q^2$--variation is much more modest. The same
happens at large $Q^2$ for the Bjorken sum rule in deep inelastic 
scattering which is free of certain assumptions entering the 
Ellis--Jaffe sum rules for
protons and neutrons. Based on the results of \cite{Ji2}, it was
argued in \cite{VB} that the chiral prediction should only be taken
seriously for $Q^2 \le 0.1\,$GeV$^2$ and that a continuous matching
between the CHPT prediction and the pQCD evolution of the Bjorken sum 
rule to order $\alpha_s^3$ can be questioned. Such type of smooth
transition from the perturbative to the non--perturbative regime
had been suggested from the study of the sum rules for nucleon
polarizabilities at finite virtuality \cite{EKW}. Note also that 
a vigorous experimental effort is underway at Jefferson Laboratory
to determine the spin structure of the proton and the neutron. 
In this letter, we report
about a novel analysis of the generalized Gerasimov--Drell--Hearn
sum rule for low photon virtualities. We employ the Lorentz--invariant
formulation of baryon chiral perturbation theory proposed in
\cite{BL}, which in contrast to the heavy 
fermion approach automatically fulfills all strictures from
analyticity (see also the earlier work in \cite{ET}).
It furthermore has the advantage of resumming recoil corrections to
all orders which in certain cases can improve the convergence
dramatically, see e.g. the discussion of the neutron electric form
factor in \cite{KM}.

\medskip
\noindent {\bf 2.}
Next, we briefly summarize the formalism necessary for the discussion
of the photon--nucleon Compton amplitude and the related sum rules.
It is common to express the spin amplitude of forward doubly virtual
Compton scattering in terms of two structure 
functions, called $S_1(\nu,Q^2)$ and $S_2(\nu,Q^2)$, via
\begin{eqnarray}
T^{[\mu\nu]}(p,q,s)=-i\;\epsilon^{\mu\nu\alpha\beta}q_\alpha\left\{m s_\beta
S_1(\nu,Q^2)+\left[p\cdot q\, s_\beta-s\cdot q\;p_\beta\right]\frac{S_2(\nu,Q^2)}{m}
\right\},
\end{eqnarray}
where $s^\mu$ denotes the spin-polarization four-vector of the
nucleon, $m$ is the nucleon mass, $\epsilon^{\mu\nu\alpha\beta}$ the
totally antisymmetric Levi--Civita tensor with $\epsilon^{0123}=1$,
$Q^2  \ge 0$ (the negative of) the
photon virtuality   and $\nu=p\cdot q/m$ the energy transfer.
Note that while $S_1 (\nu,Q^2)$ is even under crossing 
$\nu \leftrightarrow -\nu$,  $S_2(\nu,Q^2)$ is odd. In what follows,
we will be concerned with the amplitudes $\bar{S}_i (0,Q^2) =
S_i (0,Q^2) - S_i^{\rm el}  (0,Q^2)$, i.e. the Compton amplitudes
with the contribution from the elastic intermediate state subtracted.
The generalized sum rule proposed in \cite{BKM} utilizes the
imaginary part of the photon--nucleon
spin--flip amplitude $f_2 (\nu, Q^2)$, which  can be expressed in terms of
the combination $\bar{S}_1 (0,Q^2)- (Q^2/m)
d\bar{S}_2(\nu,Q^2)/d\nu|_{\nu=0}$ \footnote{Note the different 
convention as compared to Ref.~\cite{Ji1}.}. 
We concentrate here on
the sum rule \cite{Ji2}
\beq
\int_{\nu_0}^\infty \frac{d\nu}{\nu}\, G_1 (\nu, Q^2) = \frac{1}{4e^2}
\, \bar{S}_1 (0,Q^2) = {1\over m^2}\, \bar{I} (Q^2)~,
\eeq
with $G_1(\nu, Q^2)$ a spin--dependent structure function, $G_1(\nu,
Q^2)= g_1(Q^2/2m\nu , Q^2)/ (m\nu)$ and $g_1$ the dimensionless scaling
function. Furthermore, $\nu_0$ denotes the inelastic threshold.
The subtracted Compton amplitude $\bar{S}_1(\nu,Q^2)$ is normalized 
as $\bar{S}_1 (0,0) = -e^2 \kappa^2/m^2$, with $\kappa$ the anomalous magnetic
moment of the spin-1/2 particle and $e^2/4\pi = 1/137.036$ the fine structure
constant.
This formula can be converted into the first moment
of the sum rule via
\beq
\Gamma_1 (Q^2) = \frac{Q^2}{2m^2} \, \bar{I} (Q^2)~.
\eeq
All formulas given so far hold for the proton and the neutron
separately and the corresponding quantities will be indicated
by a superscript $p$ or $n$.
At small photon virtualities, the subtracted Compton amplitudes 
$\bar{S}_i (0,Q^2)$ $(i=1,2)$ can be calculated in chiral
perturbation theory, as will be discussed next. For more details
on the formalism and earlier calculations we refer to \cite{Ji1}. 

\medskip
\noindent {\bf 3.} We briefly discuss the effective chiral Lagrangian
underlying our calculation. Its generic form consists of a string
of terms with increasing chiral dimension,
\beq\label{L}
{\cal L}_{\rm eff} =
{\cal L}_{\pi N}^{(1)} + {\cal L}_{\pi N}^{(2)} + {\cal L}_{\pi N}^{(3)} +   
{\cal L}_{\pi N}^{(4)} + {\cal L}_{\pi\pi}^{(2)} + {\cal L}_{\pi\pi}^{(4)}
+ \ldots~.
\eeq
The superscript denotes the power in the genuine small parameter $q$ 
(denoting pion masses and/or external momenta). A complete one--loop (fourth
order) calculation must include all tree level graphs with insertions
from all terms given in Eq.~(\ref{L}) and loop graphs with at most
one insertion from ${\cal L}_{\pi N}^{(2)}$. The complete Lagrangian
to this order is given in \cite{FMMS}. We note that for the case under
consideration the only appearing dimension two LECs, called $c_6$ and $c_7$
\cite{BKMrev}, can be fixed from the anomalous magnetic moment of the proton and
of the neutron, respectively. 

\medskip\noindent
Baryon chiral perturbation theory 
is complicated by the fact that the nucleon mass does not vanish in
the chiral limit and thus introduces a new mass scale apart from the
one set by the quark masses. Therefore, any power of the quark masses 
can be generated by chiral loops in the nucleon (baryon) case, spoiling the
one--to--one correspondence between the loop expansion and the one
in the small parameter $q$. One method to overcome this is  the heavy mass
expansion (called HBCHPT) where the nucleon  mass is
transformed from the propagator into a string of vertices with
increasing powers of $1/m$. Then, a consistent power counting emerges.
However, this method has the disadvantage
that certain types of diagrams are at odds with strictures from analyticity.
The best example is the so--called triangle graph, which enters e.g. the
scalar form factor or the isovector electromagnetic form factors of the
nucleon.  In a fully relativistic treatment,
such constraints from analyticity are automatically fulfilled. It was recently
argued in~\cite{ET} that relativistic one--loop integrals can be separated
into ``soft''' and ``hard'' parts. While for the former the power counting
as in HBCHPT applies, the contributions from the latter can be absorbed in 
certain  low--energy constants (LECs). In this way,
one can combine the advantages 
of both methods. A more formal and rigorous implementation of such a program 
was given in \cite{BL}. The underlying method  is called 
``infrared regularization''. Any dimensionally regularized
one--loop integral $H$ is split into an infrared singular and a regular part
by a particular choice of Feynman parameterization. Consider first the
regular part, called $R$. If one chirally  expands these terms, one generates
polynomials in momenta and quark masses. Consequently, to any order, $R$ can
be absorbed in the LECs of the effective Lagrangian.  On the other hand, the
infrared (IR) singular part $I$ has the same analytical properties as the full
integral $H$ in the low--energy region and its chiral expansion leads to the
non--trivial momentum and quark--mass dependences of CHPT, like e.g. the
chiral logs or fractional powers of the quark masses.
For a typical one--loop integral (like e.g. the nucleon self--energy)
this splitting can be achieved in the following way
\beq
H = \int  \frac{d^dk}{(2\pi)^d} {1 \over AB} 
= \int_0^1 dz \int  \frac{d^dk}{(2\pi)^d} {1 \over [(1-z)A+zB]^2}
= \biggl\{ \int_0^\infty - \int_1^\infty \biggr\} dz
 \int  \frac{d^dk}{(2\pi)^d} {1 \over [(1-z)A+zB]^2} = I + R~,
\eeq
with $A=M_\pi^2-k^2-i\epsilon$, $B=m^2 -(p-k)^2 -i\epsilon$,
$\epsilon \to 0^+$ and $M_\pi$ the pion mass. Any general
one--loop diagram with arbitrary many insertions from external sources
can be brought into this form by combining the propagators to a single
pion and a single nucleon propagator. It was also shown
that this procedure leads to a unique, i.e.
process--independent result, in accordance with the chiral Ward
identities of QCD \cite{BL}.  
Consequently, the transition from any one--loop graph $H$
to its IR singular piece $I$ defines a symmetry--preserving regularization.

\medskip\noindent
Within this approach, we have calculated $\bar{S}_{1,2}^{(p,n)} (0,Q^2)$. 
The pertinent one--loop diagrams are given in \cite{Ji2}, however,
in the approach used here one does not need to consider 
dimension two insertions from the kinetic energy on the nucleon
lines since this is done automatically in a Lorentz--invariant
formulation. In fact, all such diagrams are counting as third order
in the approach used here, the only genuine fourth order graphs are
the ones with one insertion $\sim c_{6,7}$ (anomalous magnetic moment).
Therefore, $\bar{S}_1(\nu,Q^2)$ is already non--vanishing at third order,
quite in contrast to the heavy baryon calculation.  As a check, we remark 
that in the limit of a very large nucleon mass, one recovers the 
earlier HBCHPT 
results of \cite{BKM} and \cite{Ji2}. However, as already found in the 
relativistic calculation
of the slope of the generalized GDH sum rule in \cite{BKM}, 
one can not give the
results in closed analytical form. We refrain from giving the expressions based
on scalar loop functions here but rather refer to the upcoming 
paper \cite{BHM}. Of course, there are also non--negligible  
resonance contributions to the $\bar{S}_i (\nu,Q^2)$. More precisely,
the effects from the $\Delta$ largely cancel in $\bar{S}_1 (\nu,Q^2)$
\cite{Ji2} but are expected to be more important in $\bar{S}_2 (\nu,Q^2)$. There are
additional (smaller) vector meson and higher baryon resonance contributions.
Therefore  we mostly concentrate here on $\bar{S}_1 (\nu,Q^2)$ 
and also on the neutron--proton difference
in which most resonance  effects cancel, see e.g. \cite{BI,BuL}.

\begin{figure}[t]
\parbox{.48\textwidth}{\epsfig{file= 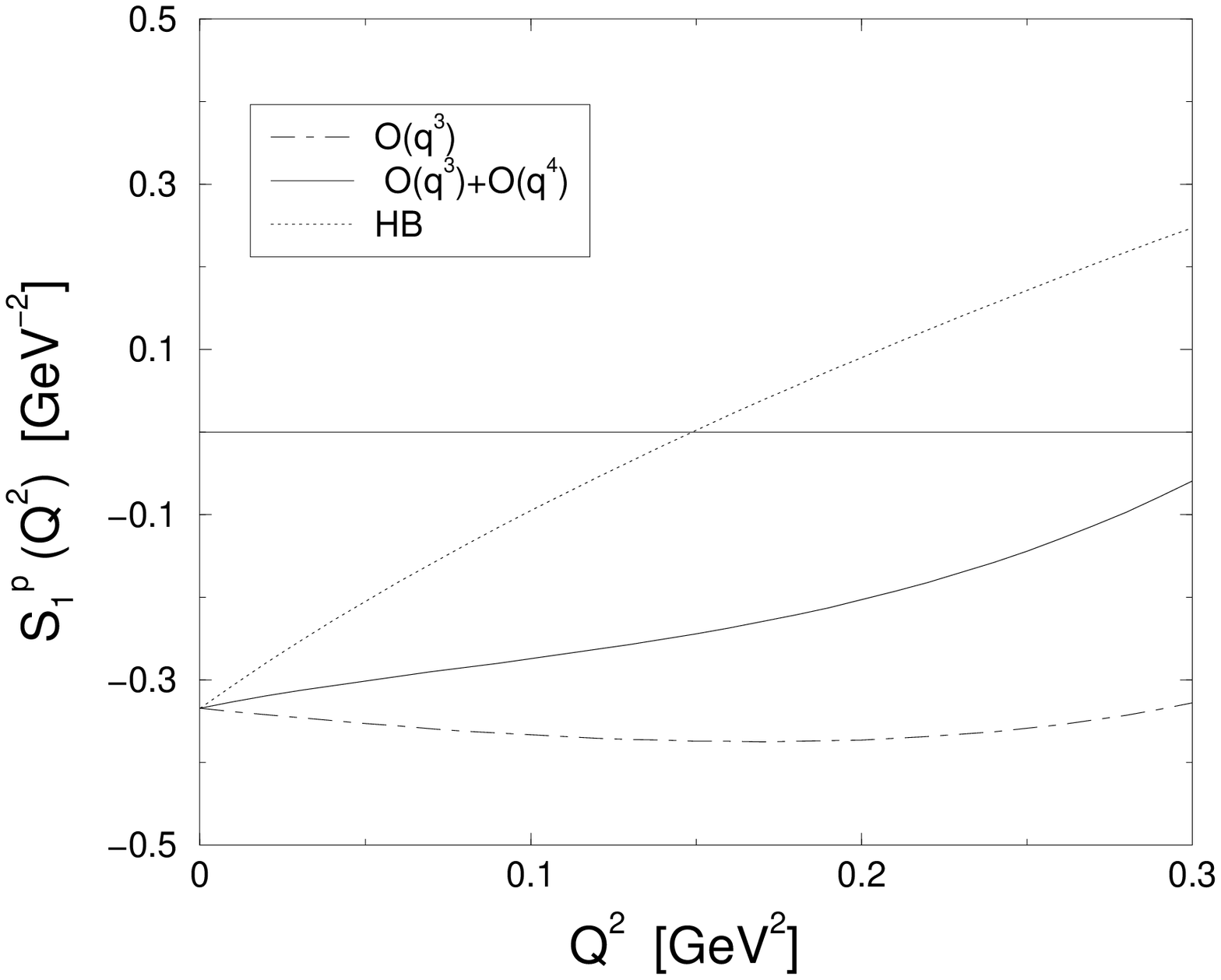,width=.44\textwidth,silent=,clip=}}
\hfill
\parbox{.48\textwidth}{\epsfig{file= 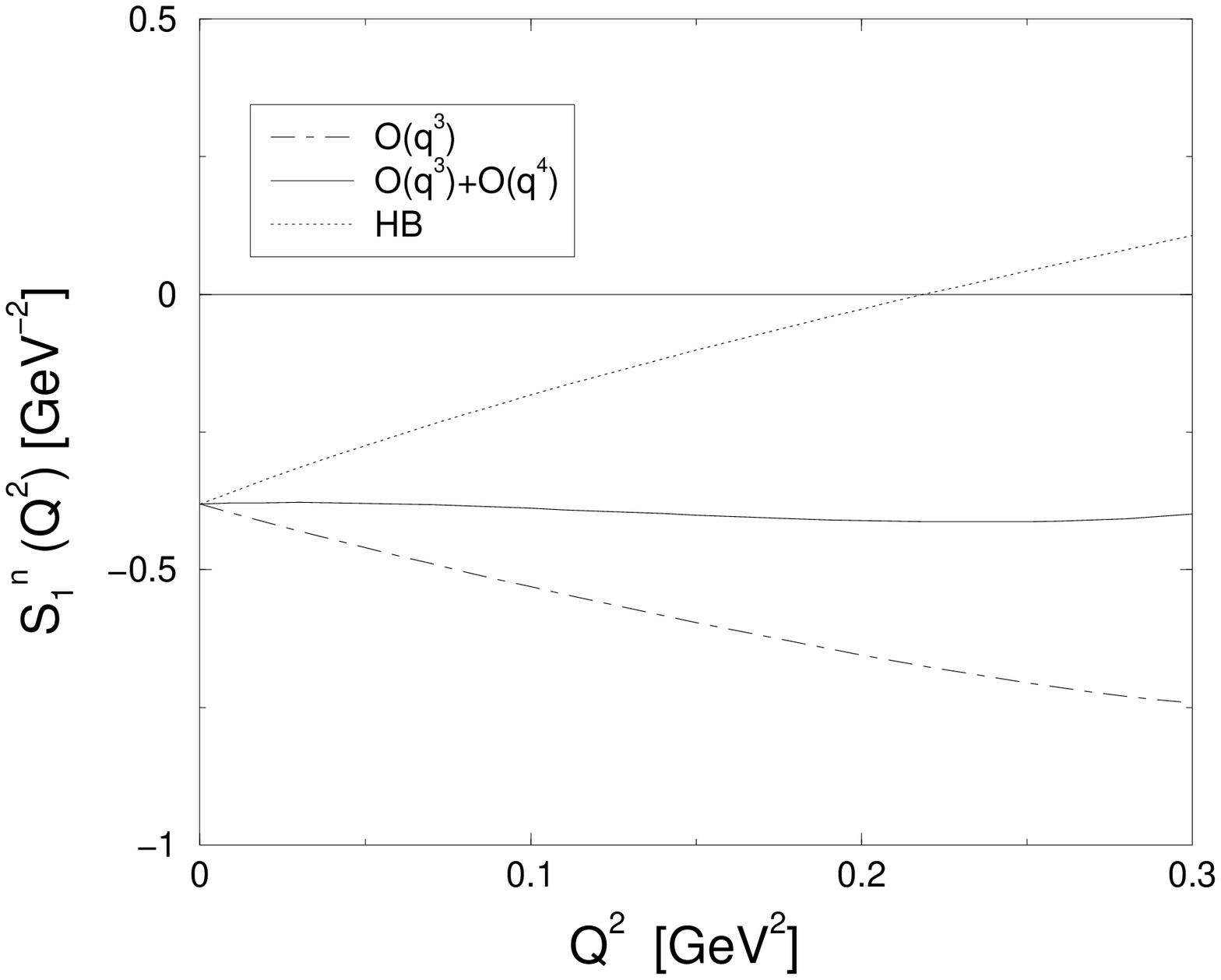,width=.44\textwidth,silent=,clip=}}
\caption{\label{fig:s1pn}
          Chiral loop contribution to the structure function $\bar{S}_1
          (0,Q^2)$ with the elastic contribution subtracted. 
          The solid (dot-dashed) line gives the result of
          the present calculation to order $q^4$ ($q^3)$ in comparison
          to the heavy baryon result of \protect\cite{Ji2} (dotted line).
          Left (right) panel: Proton (neutron).
}
\end{figure}

\medskip
\noindent {\bf 4.}
We now present our results. First, we must specify the parameters.
We use $g_A = 1.267$, $F_\pi = 93\,$MeV, $M_\pi = 139.57\,$MeV, $m =
0.9389\,$MeV, $\kappa_v =3.706$ and $\kappa_s = -0.120$. The latter
two quantities include the LECs $c_6$ and $c_7$. 
We point out again that there are no unknown parameters and thus
we obtain genuine one--loop predictions. In Fig.~\ref{fig:s1pn}
we show the results for $\bar{S}_1^{p,n} (0,Q^2)$ in comparison to the heavy
baryon result of \cite{Ji2}. Some remarks are in order. First, we note
that in the IR approach, one has third and fourth order contributions with
the latter becoming sizeable for $Q^2 \ge 0.2\,(0.3)$GeV$^2$ for the
proton (neutron) \footnote{We remark that for certain tree graphs
at $\nu =0$ the heavy baryon propagator formally scales as ${\cal O}(p^{-2})$ so
that one could consider the contribution generated from such diagrams
also as third order in HBCHPT, for details see \cite{BHM}}. Second, the
modulus of the structure function is bigger as compared to the HBCHPT
result. Similar remarks apply to the results for $\bar{S}_2^{p,n} (0,Q^2)$
not shown here. These sizeable differences between our results and the
HBCHPT ones can be traced back to the fact that the $1/m$ expansion for
these structure functions is very slowly converging, and thus the method
used here appears to be more appropriate.
\noindent
{}From these results one deduces the following 
slopes for the proton and neutron
structure function $\bar S_1 (Q^2)$ (from here on, we do not display
the argument $\nu = 0$ any longer): 
\beqa
\frac{d\bar{S}_1^p (Q^2)}{dQ^2}\Bigl|_{Q^2=0} &=& 
(-0.4 + 1.2)~{\rm GeV}^{-4} = 0.8~{\rm GeV}^{-4}~,
\\
\frac{d\bar{S}_1^n (Q^2)}{dQ^2}\Bigl|_{Q^2=0} &=&
(-1.7 + 1.9)~{\rm GeV}^{-4} = 0.2~{\rm GeV}^{-4}~,
\eeqa
where the first (second) term in the round brackets refers to the
third (fourth) order contribution. These results are
compatible with expectations from naive dimensional analysis. 
For comparison, the corresponding
HBCHPT results are $d\bar{S}_1^p / dQ^2 = 2.9\,$GeV$^{-4}$ and
$d\bar{S}_1^n / dQ^2 = 2.4\,$GeV$^{-4}$, which are quite different.
The large slope in the heavy baryon case is due to a variety of effects.
First, the large value of the isovector anomalous magnetic moment, 
$\kappa_v = 3.71$, appearing in
$d\bar{S}_1/dQ^2\propto [1+3\kappa_v+2(1+3\kappa_s)\tau_3]$ leads
to a prefactor for the proton of 14 instead of the
expected order of one based on naive dimensional analysis.
Second, as as been spelled out clearly in \cite{GHM}, in the spin
sector the natural expansion parameter seems to be $\pi M_\pi/m$
rather than  $M_\pi/m$ (for the odd powers), 
which can lead to sizeable corrections.
Consequently, the $1/m$ expansion converges slowly, and thus
the numbers for these slopes found here should be considered
more reliable. 
Note also that for these  slopes the relative size of the fourth
order corrections comes out very different for the proton and the
neutron, but it should be stressed that
the meaningful numbers are the {\em full} one--loop (third and fourth order)
results. For orientation, we 
quote the result of the dispersive analysis of \cite{DKT}, which
besides pion loop effects also  includes resonance contributions, 
it is $d\bar{S}_1^p / dQ^2 = 1.8\,$GeV$^{-4}$.

\begin{figure}[t]
\parbox{.48\textwidth}{\epsfig{file= 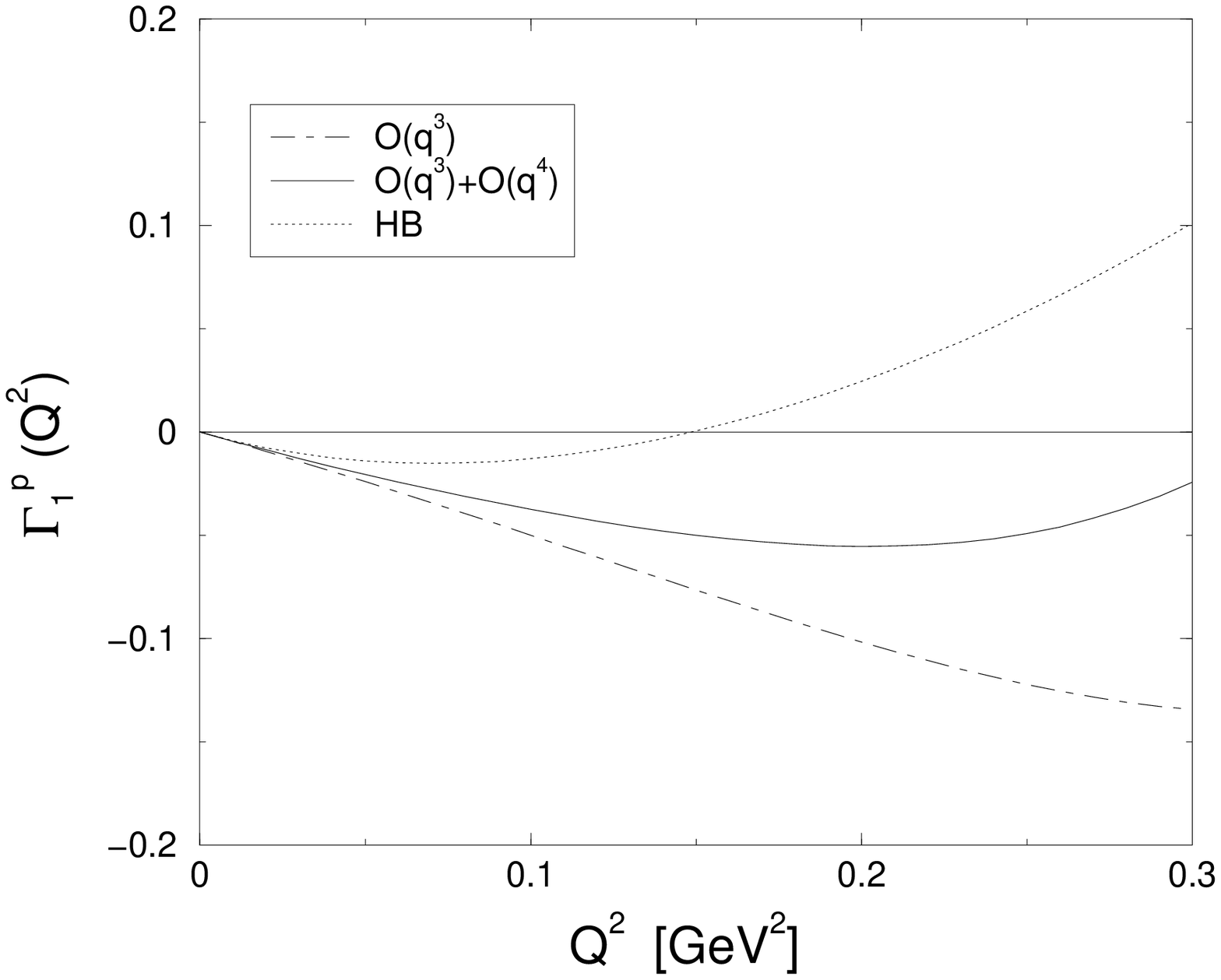,width=.44\textwidth,silent=,clip=}}
\hfill
\parbox{.48\textwidth}{\epsfig{file= 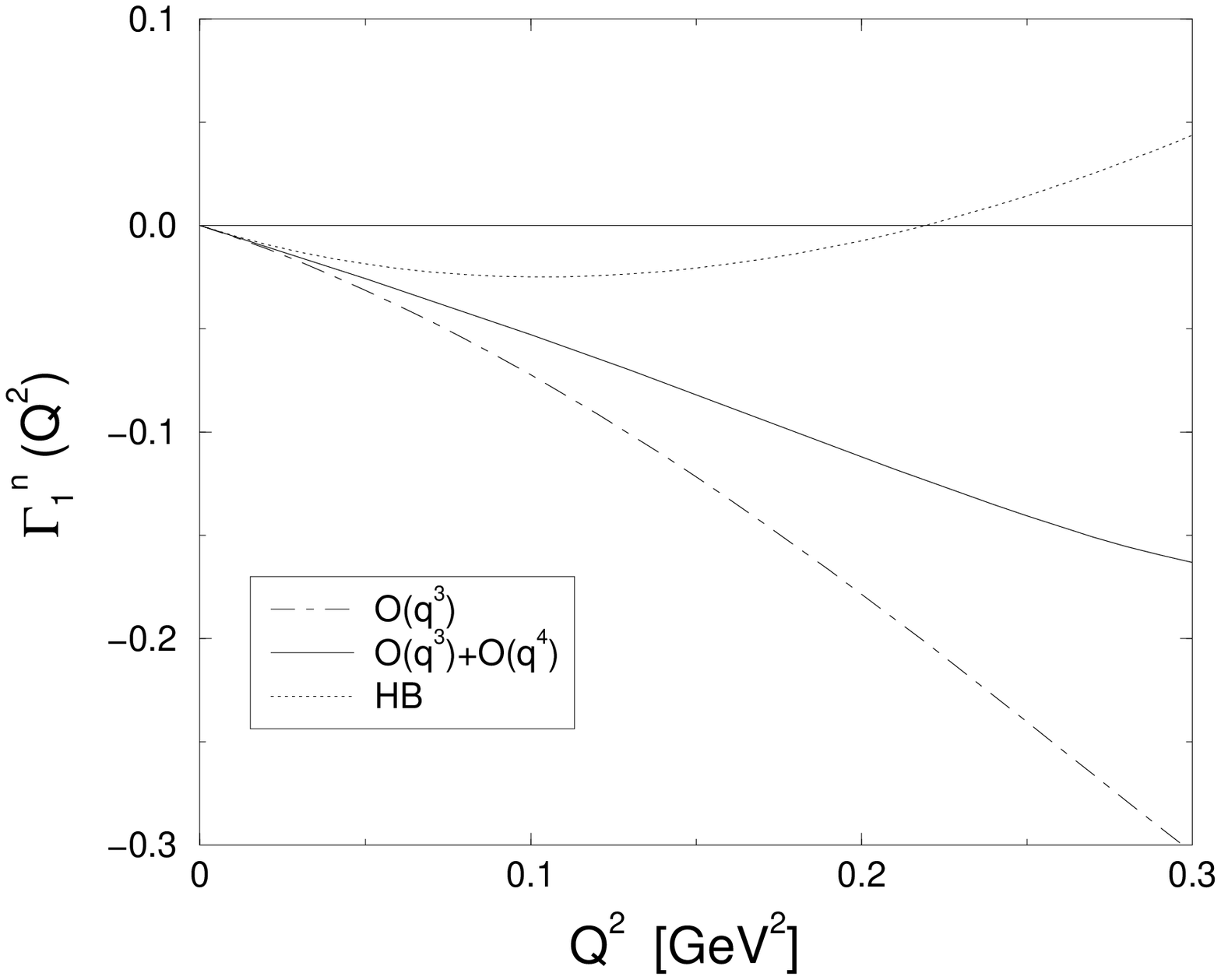,width=.44\textwidth,silent=,clip=}}
\caption{\label{fig:ga1pan}
          Chiral loop contribution to the first moment $\Gamma_1
          (Q^2)$. The solid (dot-dashed) line gives the result of
          the present calculation to order $q^3$ ($q^4)$ in comparison
          to the heavy baryon result of \protect\cite{Ji2} (dotted line).
          Left (right) panel: Proton (neutron).
}
\end{figure}

\medskip\noindent
In Fig.~\ref{fig:ga1pan} we show the first moments $\Gamma_1^p (Q^2)$ and
$\Gamma_1^n (Q^2)$ in comparison to the heavy baryon results. 
For the range of photon virtualities below $Q^2 = 0.2\,(0.3)\,$GeV$^2$, 
the fourth order corrections are not dramatically
large for the proton (neutron). Note that while $\Gamma_1^p (Q^2)$ crosses
zero at $Q^2 \simeq 0.33\,$GeV$^2$, $\Gamma_1^n (Q^2)$ stays negative for 
all values of $Q^2$ in the range of virtualities considered here.
We stress again that for a 
direct comparison with the data, some resonance contributions have to be added, 
this will be done and discussed in the upcoming work \cite{BHM}. However,
one does not expect these resonance contributions to be overly large.
\noindent
Note that we refrain from comparing these curves with the preliminary data
from Jefferson Laboratory because these data are not yet
final and can therefore not be shown. This comparison will be done
in \cite{BHM}.

\begin{figure}[htb]
\centerline{
\epsfysize=2.72in
\epsffile{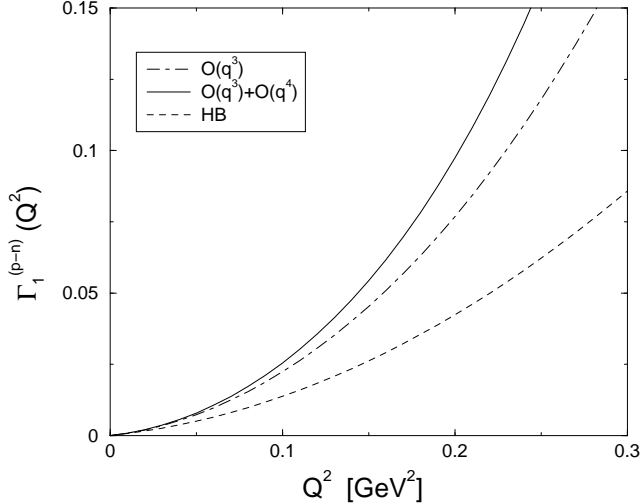}
}
\begin{center}
\parbox{10cm}{\caption{\protect First moment difference $\Gamma_1^p
    -\Gamma_1^n$. The solid and the dot-dashed line stand for the chiral
   expansion at fourth  and third order, respectively, using the 
   Lorentz--invariant approach and the dotted line represents the result of the
   heavy baryon formalism \cite{Ji2,VB}.
\label{fig:ga1pn}}}
\end{center}
\end{figure}

\medskip\noindent
In Fig.~\ref{fig:ga1pn}
we show the result for $\Gamma_1^{p-n} (Q^2)$ in comparison to the 
HBCHPT result \cite{VB}. This should be considered the main result of
this paper. For $Q^2 \leq 0.1\,$GeV$^2$, the fourth order IR and the
heavy baryon result are very similar. For larger momentum transfer,
however, $\Gamma_1^{p-n} (Q^2)$ grows faster with
increasing momentum compared to the heavy baryon calculation. This is due to
the fact that the chiral corrections to the individual $\Gamma_1^p (Q^2)$ 
and $\Gamma_1^n (Q^2)$ are quite different (similar) in the IR (HB)
approach (for $Q^2 \geq 0.1\,$GeV$^2$).
The corresponding slope is of course given by the GDH sum rule,
\beq
\frac{d\Gamma_1^{p-n} (Q^2)}{dQ^2}\Bigl|_{Q^2=0} =
\frac{\kappa_n^2 - \kappa_p^2}{8m^2} = 0.063~{\rm GeV}^{-2}~,
\eeq
as can also been seen in the figure since all curves merge to one
around $Q^2 \simeq 0$.

\medskip\noindent
It is also interesting to consider the 
slopes corresponding to the combination 
$\bar{S}_{12} (Q^2) \equiv \bar{S}_1 (0,Q^2)
-(Q^2/m)$ $\times (d\bar{S}_2(\nu,Q^2)/d\nu)|_{\nu =0}$
for the proton and the neutron case separately, we find
\beqa
\frac{d\bar{S}_{12}^p (Q^2)}{dQ^2}\Bigl|_{Q^2=0} 
&=&  (3.2 + 1.8)~{\rm GeV}^{-4} = 5.0~{\rm GeV}^{-4}~,
\\
\frac{d\bar{S}_{12}^n (Q^2)}{dQ^2}\Bigl|_{Q^2=0} 
&=&  (-0.3 - 1.8)~{\rm GeV}^{-4} = -2.1~{\rm GeV}^{-4}~,
\eeqa
which should be compared with the heavy baryon results \cite{BKM,Ji2},
$d\bar{S}_{12}^p (Q^2)/dQ^2 = 
3.7\,(1 -2.6)~{\rm GeV}^{-4} = -6.0~{\rm GeV}^{-4}$ and
$d\bar{S}_{12}^n (Q^2) /dQ^2 = 
3.7\,(1 -2.1)~{\rm GeV}^{-4} = -4.3~{\rm GeV}^{-4}$. 
We remark that the seemingly overwhelming fourth order correction for the
neutron is simply a reflection of the accidentally suppressed third order
result. The 50\% correction found for the proton is not uncommon
for the nucleon spin sector. 
The larger corrections in HBCHPT are simply
a reflection of the various effects discussed already before,
In the IR approach used here, higher order $1/m$ 
corrections (in the HBCHPT counting) tend to diminish such large corrections.
We point out that such effects also appear in other observables, like
e.g. in pion photoproduction off nucleons, but in that case their effect is
much less dramatic because they are accompanied by small or normal prefactors
(less than or of order one) or do not play a significant role. Such large
corrections are only appearing in the spin sector and their origin is essentially
understood. 
Note further that the
slope of the integrals $I_{p,n} (0)$ defined in \cite{BKM} differ
from the conventions used here by an overall factor of $\pi/2$.
For this combination of the structure functions, one has to include
the $\Delta$ (and other resonance)  contribution for a direct comparison 
with the data,
see e.g. \cite{BKM}.

\medskip
\noindent {\bf 5.} In this paper, we have considered
the one--loop representation of the structure functions of doubly virtual
Compton scattering at low photon virtualities,
$\bar{S}_i (0,Q^2)$, $i=1,2$ (with the elastic intermediate state
contribution subtracted). We have performed the
analysis in the framework of infrared regularized relativistic baryon
chiral perturbation theory and compared the results with the existing
ones obtained in the heavy mass formalism. We have found that there
are significant differences for the individual structure functions
$S_1^p (0,Q^2)$ and $S_1^n (0,Q^2)$, i.e. these exhibit a less
dramatic (stronger) $Q^2$--evolution for the proton (neutron) 
than in HBCHPT and show reasonable
convergence for small photon virtualities. We have also pointed out that
the $1/m$ expansion underlying the heavy baryon approach is only
slowly converging for these structure functions and thus the resummation
of recoil effects inherent to the IR formalism is to be considered important.
Clearly, these predictions should only be compared to data when the
missing resonance contributions are included. 
However, in the first moment difference 
$\Gamma_1^{p-n} (Q^2)$, contributions from spin--3/2 resonances like the
delta drop out (and most other resonance contributions, too), so that 
the predictions shown in Fig.\ref{fig:ga1pn} should be considered the main
result of this paper.  As it turns out, the growth of $\Gamma_1^{p-n} (Q^2)$ 
with increasing $Q^2$ is stronger to what was found previously  
in the heavy mass formulation. It would be interesting to compare this
prediction with the data soon becoming available from Jefferson
Laboratory\footnote{First preliminary data from the CLAS
collaboration which start at $Q^2 = 0.2\,$GeV$^2$ seem to be consistent
with our and the heavy baryon result \cite{RdV}.}. It is important
to systematically add resonance contributions to these chiral predictions
for the proton and the neutron separately
to allow for a direct comparison with the data. Work along these lines is
underway \cite{BHM}.

\bigskip

\noindent{\bf Acknowledgments}

\medskip\noindent
We thank Volker Burkert for some useful communications and 
Norbert Kaiser for comments. One of us
(UGM) thanks Susan Gardner and Wolfgang Korsch for their hospitality
at the University of Kentucky, where part of this work was done.

\vspace{1cm}

\noindent{\bf Note added}

\medskip\noindent
After submission of this manuscript, new experimental information on the
integral $I_n (Q^2)$ for $Q^2 = 0.1 - 0.9\,$GeV$^2$ appeared (M. Amarian
et al. [The Jefferson Lab E94010 Coll.], arXiv:nucl-ex/0205020). The point
at lowest photon virtuality is in good agreement with the prediction given
here, but the prediction at $Q^2 = 0.26\,$GeV$^2$ is much more negative than
the data.


\end{document}